\documentclass[referee]{aa} 
\usepackage{graphicx}
\usepackage{txfonts}
\begin{document}
\title{Constraints on the cosmic equation of state: age conflict versus phantom energy}

\subtitle{Age--redshift relations in an accelerated universe}

\titlerunning{Age conflict versus phantom energy}

\author{J. Cepa
\inst{1,2}
}


\institute{Instituto de Astrof\'\i sica de Canarias E-38200 La Laguna, Tenerife, Spain
\and
Departamento de Astrof\'\i sica, Facultad de F\'\i sica, Universidad de la Laguna. 
E-38071 La Laguna, Tenerife, Spain\\
\email{jcn@ll.iac.es}
}

\date{Accepted March 24, 2004}

\abstract{
The currently available values and confidence limits for $\Omega_{\mathrm{m0}}$, $H_0$ and 
globular cluster ages still indicate that the dark energy that dominates the Universe 
could also be a form of quintessence or phantom energy. In fact, current cosmological values 
favor phantom energy. To increase the likelihood of a cosmological constant as dark energy
instead of phantom energy, the possibilities seem to lie in reducing globular cluster ages, 
the Hubble constant, or both, and possibly advancing the epoch of globular cluster formation. 
For a set of possible dark energy equations of state that includes the cosmological constant, 
quintessence or phantom energy, age--redshift analytical expressions for null curvature 
universes that include ordinary matter are derived together with the corresponding ages 
for these universes. ``Cosmic coincidences'' are found for the currently accepted values of 
$\Omega_{\mathrm{m0}}$ when evaluating the age of the Universe with a cosmological constant 
or a phantom energy as dark energies.
}

\maketitle


\section{Introduction}

Curvature, age, the Hubble constant and energy density form a long-standing cosmological 
puzzle. Measurements of the cosmic microwave background radiation (CMBR) anisotropy have confirmed 
that the Universe is consistent with null, or nearly null, curvature (de Bernardis et al.\ 
\cite{bernardis}), as predicted by the inflation paradigm. However, a Euclidean 
universe composed of ordinary matter only has an age inconsistent with that derived from 
globular clusters (Bolte \& Hogan \cite{bolte}). Moreover, galaxy cluster data 
(Carlberg et al.\ \cite{carlberg}) allow us to obtain only about 20--30\% of the amount of 
matter required for achieving null curvature. This puzzle seemed to be solved by the discovery 
in 1998 of the first evidence for an accelerating universe based on distant type Ia 
Supernova (SNeIa) observations (Riess et al.\ \cite{riess1}), which have been supported by 
additional data including more distant SNe (Riess et al.\ \cite{riess2}). This acceleration
implies the presence of a significant fraction of dark energy with an equation of state 
of negative pressure. The addition of the required amount of this energy to the 
observed amount of matter allows us to obtain the total amount of energy required for 
achieving null curvature. It  also has the desired side effect of producing an older 
universe that can then be made consistent with independent age measurements.

However, the nature of this dark energy is still to be established. Moreover, its 
equation of state is still unknown. The first attempts rescued the cosmological 
constant as an ad hoc explanation, although their interpretation of the constant in terms of a 
vacuum energy is inconsistent by 124 orders of magnitude with respect to the required 
value, which leads us to assume that its value has to be zero for consistency. Other 
energy types, such as quintessence or phantom energy, have also been invoked, although their 
origin is no clearer than that of the cosmological constant. 

Quintessences were introduced by Wetterich (\cite{wetterich}), Caldwell et al. 
(\cite{caldwell1}), Ratra \& Peebles (\cite{ratra}) to avoid the extreme fine-tuning 
needed to allow a cosmological constant to be significant only at recent epochs. 
Quintessences are characterized by a scalar field whose evolution depends on its 
potential. The most popular potentials are inverse power  and exponential laws 
(Kneller \& Strigari \cite{kneller} and references therein), although other possibilities
have been considered (see, for example, Di Pietro \& Claeskens\ \cite{dpc}). Both inverse power
laws and exponential potentials converge to a unique solution for a broad range of different 
initial conditions.

Phanton energy (Caldwell \cite{caldwell2}) violates 
the dominant-energy condition (Hawking \& Ellis\ \cite{he}) that might allow the 
existence of wormholes. It also makes the Universe reach in a finite time a cosmic doomsday 
where all objects, from galaxies to nucleons, are ripped apart, a situation that has been
termed the ``big rip'' (Caldwell \cite{caldwell2}). Although some Chaplygin gas model generalization
can avoid this big rip (Gonz\'alez-D\'\i az \cite{gd}), the violation of the dominant energy 
condition and the difficulties in obtaining a stable phantom model (Carroll et al.\ 
\cite{carroll}) renders this kind of energy more problematic  than quintessences. 

In the case of either quintessences, the cosmological constant or phantom energy, it seems 
established that the introduction of negative pressure energies is required to fit and 
harmonize the existing CMBR anisotropy, SNeIa, clusters of galaxies, large scale structure, 
Big Bang nucleosynthesis and age estimator data.

The incorporation of these energies changes the age--redshift relation and estimates 
of the age of the Universe. The aim of this article is to show the difficulty in avoiding 
phantom energies with the current age estimator data and to provide analytical age expressions 
for the homogeneous and isotropic case for the most likely equations of state for 
our Universe, considered constant, as restricted by the cosmological parameters 
commonly assumed. These include the cosmological constant, quintessence and phantom energy.

Thomas \& Kantowski (\cite{thomas}) provide an age--redshift relation using elliptical 
integrals for a universe composed of matter and a cosmological constant, but without 
providing an easy ready-to-use expression. In general, no analytical expressions 
of age as a function of redshift seems to be readily available in the literature for the 
currently most widely accepted cosmologies: flat universes of matter plus a cosmological 
constant, quintessence or phantom energy. 

Although numerical integration allows us to obtain ages in a relatively straightforward 
way (see, for example, Hogg \cite{hogg}), analytical solutions are faster to evaluate and
have the advantage of providing more precise and ready results than numerical approaches. 
Moreover, analytical expressions provide the explicit dependence on energy densities, 
thereby easing their study. Finally, they constitute a reference for checking numerical 
solutions. 

In Section 2 the basic equations are given. In Section 3 possible values for the 
equation of state, considered constant, are reviewed according to recent figures for 
the cosmological parameters, in Section 4 the general case of a non-constant equation 
of state is outlined, and in Section 5 analytical age--redshift relations and ages are 
obtained by solving the equations on a case by case basis for the range of the constant 
equations of state assumed. 


\section{Basic equations}

The age--redshift relation can be obtained from the Friedmann equations. Using  units 
defined by $c=G=1$, where $c$ is the speed of light and $G$ the constant of gravitation,
the Friedmann equation representing energy balance can be written as

\begin{equation}
H^2=\left(\frac{\dot a}{a}\right)^2=\frac{8\pi}{3}\rho-\frac{\epsilon}{a^2},
\label{friedmann}
\end{equation}
where $a$ is the scale factor at a cosmological time $t$, $\rho$ the total energy density, 
$\epsilon$ the curvature parameter and the dot indicates the time derivatives. 

The total energy density as a function of redshift, $z$, is defined to be

\begin{equation}
\frac{8\pi}{3}\rho= H_0^2 \left(\Omega_{\mathrm{m0}}(1+z)^3+
\Omega_{\mathrm{w0}}(1+z)^{3(1+w)}\right),
\label{ro}
\end{equation}
where subscript 0 indicate present-day magnitudes of time-dependent functions,
$\Omega_{\mathrm{m0}}$ and $\Omega_{\mathrm{w0}}$ are the present-day energy densities
over the critical density of ordinary matter and of dark energy, respectively, 
$1+z=a_0/a$, and $w=p_w/\rho_w$ is the equation of state of dark energy given 
by the ratio of scalar pressure over energy density, which is assumed constant. 

This assumption, a generalization of the cosmological constant that 
represents a barotropic fluid, lacks physical justification. A constant $w$ is valid 
for the cosmological constant only, since in general a time dependency is expected 
in quintessences and phantom energies. However, the absence of well motivated dark energy 
models based on fundamental physics and the difficulty to observe a time dependence of $w$ 
either by using the CMBR (Aurich \& Steiner \cite{aurich}) or fits to luminosity distances 
(Di Pietro \& Claeskens\ \cite{dpc}), makes this assumption a reasonable model-independent 
approximation to the equation of state of these energies (see Kneller \& Strigari 
\cite{kneller} and references therein for advantages and limitations of this approach). 
As a consequence, the constant $w$ values inferred represent effectives values 
weighted with respect to $\Omega_{\mathrm{w0}}$ (Doran et al. \cite{doran1}; 
Doran \& Lilley \cite{doran2}). Nevertheless, a more general case is briefly outlined
in Section 4.

For an $\epsilon=0$ universe, then,  the age--redshift relation can be obtained using

\begin{equation}
t=\frac{1}{H_0}\int_{z}^{\infty}\frac{\mathrm{d}z}{(1+z)\sqrt{\Omega_{\mathrm{m0}}(1+z)^3+
\Omega_{\mathrm{w0}}(1+z)^{3(1+w)}}},
\label{laecuacion}
\end{equation}
where $-1/3<w<-1$ for quintessences, $w=-1$ for the cosmological constant and $w<-1$ 
for phantom energies.

The age of the Universe, $t_0$, for a given combination of $\Omega_{\mathrm{m0}}$, 
$\Omega_{\mathrm{w0}}$ and the Hubble constant, $H_0$, can be obtained by setting
the lower limit of (\ref{laecuacion}) to $z=0$.


\section{Constraints on the cosmic equation of state}

Fits of observational data of good standard candles to distance--redshift relations can 
provide limits on the parameter $w$ of the equation of state. The main limitations of this
method comes from the uncertainties in the extinction and evolution of the selected candles 
and in the curvature, $H_0$ and energy densities. Constraining the 
dark energy equation of state from the age of the Universe requires an accurate knowledge 
of $H_0$, $\Omega_{\mathrm{m0}}$, the curvature and $t_0$, where $t_0$ is obtained from the age, $t_{\mathrm{GC}}$, 
of globular cluster or halo stars used as age indicators, 
plus the time, $t_{\mathrm{f}}$, elapsed from the Big Bang to globular cluster formation. 
For this article, the following parameters have been assumed:

\begin{enumerate}
\item A flat Universe ($\epsilon=0$) as inferred from CMBR data. This result does not 
depend on the cosmological model (see for example Hu et al.\ \cite{hu} ; Knox et al.\ \cite{knox}). 
\item The Hubble constant $H_0=72\pm 8$ km\ s$^{-1}$\ kpc$^{-1}$ as obtained by the HST Key Project to Measure the 
Hubble Constant (Freedman et al.\ \cite{freedman}), since it is derived from local Universe objects, 
whose distances are not measurably affected by the cosmological model assumed.
\item The density of matter can be obtained from the multiple source data statistical 
analysis of Turner (\cite{turner}), which give $\Omega_{\mathrm{m0}}=0.33\pm 0.035$ (68\% c.l.). 
\item The age of the Universe is obtained from the age, $t_{\mathrm{GC}}$, of galactic globular 
clusters as 12.9 $\pm$ 2.9 Gyr at 95\% c.l. using Hipparcos astrometric data 
(Carretta et al.\ \cite{carretta}). From HST data the same authors estimate the 
epoch, $t_\mathrm{f}$, of formation of the galactic globular clusters to be $z=3$, which approximately corresponds 
to 2 Gyr depending on cosmology. This is the same $t_\mathrm{f}$ as assumed by 
Spergel et al.\ (\cite{spergel}). Hence, $10.0+t_\mathrm{f}$ Gyr represents an estimate for the 
minimum age of the Universe, where $t_\mathrm{f}$ depends on the equation of state through the 
age--redshift relation. 
\end{enumerate}

\begin{figure}
\resizebox{\hsize}{!}{\includegraphics{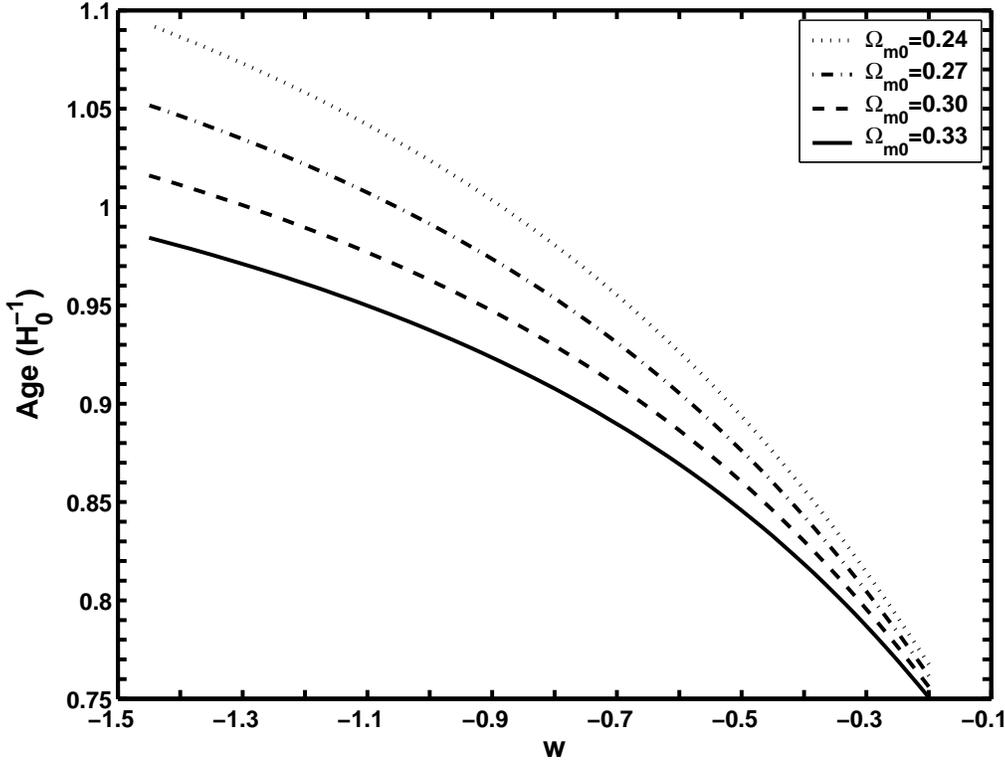}}
\caption{Universe age \emph{versus\/} the equation of state of dark energy, $w$, for a null curvature
universe with ordinary matter. The age becomes less sensitive to $w$ as $w$ decreases.
}
\label{Figure1}
\end{figure}

From each of the $\Omega_{\mathrm{m0}}$ lower limits derived from the 68\% and 95\% c.l.  
of this parameter,
values of the equation of state, $w$, are obtained by numerical integration of (\ref{laecuacion}) 
assuming null curvature and consistency with the lower limits for the ages of globular clusters
derived from their 68\% and 95\% c.l. These values constitute upper limits for $w$ because smaller 
values would be consistent with a larger $t_0$ and/or larger $\Omega_{\mathrm{m0}}$. Since 
$t_\mathrm{f}$ corresponds to $z=3$, an iterative procedure has been followed. 
Given that most of the uncertainties in the Hubble constant are systematic 
(Freedman et al.\ \cite{freedman}), two $H_0$ values, 
corresponding to 
the mean and to the lowest boundary, have been considered. The latter again yields an upper limit for $w$
since lowering $H_0$ requires increasing $w$ for the same age and $\Omega_{\mathrm{m0}}$. This 
procedure allows us to obtain the results of Table \ref{table1} for the upper limits of $w$ 
for the different combination of 68\% and 95\% c.l. of $\Omega_{\mathrm{m0}}$, $t_0$ and $H_0$. 
Lower limits in $w$ have not been derived since more accurate age estimates are required because 
the age becomes less sensitive to $w$ when $w$ decreases (Figure \ref{Figure1}). 

\begin{table}
\begin{center}
\caption[]{Upper limits for $w$ obtained for the assumed values of $\Omega_{\mathrm{m0}}$ and 
$t_{\mathrm{GC}}$ at 68\% and 95\% c.l. for globular cluster formation at $z=3$.}
\label{table1}
\begin{tabular}{ccccc}
\hline
\noalign{\smallskip}
$t_0$ (Gyr) & $\Omega_{\mathrm{m0}}$ & $t_{\mathrm{GC}}$ (Gyr) & $w$ & $H_0$ (km\ s$^{-1}$\ kpc$^{-1}$) \\
\noalign{\smallskip}
\hline
\noalign{\smallskip}
13.53	& 0.295 & 11.45 & $-$1.21 & 72 \\
11.98	& 0.295 & 10.00 & $-$0.56 & 72 \\
13.66	& 0.260 & 11.45 & $-$1.02 & 72 \\
12.06	& 0.260 & 10.00 & $-$0.52 & 72 \\
13.73	& 0.295 & 11.45 & $-$0.83 & 64 \\
11.88	& 0.295 & 10.00 & $-$0.25 & 64 \\
13.82	& 0.260 & 11.45 & $-$0.57 & 64 \\
11.92	& 0.260 & 10.00 & $-$0.23 & 64 \\
\noalign{\smallskip}
\hline
\end{tabular}
\end{center}
\end{table}

From the fit to distance--redshift relations using gravitational lenses and SN data, 
Waga \& Miceli (\cite{waga}) derive $w<-0.7$ (68\% c.l.) and $w<-0.6$ (95.4\% c.l.) for 
$\Omega_{\mathrm{m0}}=0.3$. Perlmutter et al.\ (\cite{perlmutter}) obtained $w<-0.6$ (95\% c.l.) 
using SNeIa and large scale structure data. The age of the Universe was the procedure 
employed by Krauss \& Chaboyer (\cite{krauss}) from globular cluster age estimates to 
obtain $w<-0.7$ (68\% c.l.) and $w<-0.45$ (95\% c.l.). A statistical analysis from 
WMAP (Wilkinson Microwave Anisotropy Probe) CMBR data combined with SNeIa and large scale 
structure data gives $w<-0.78$ at 95\% c.l. (Spergel et al. \cite{spergel}), with the 
previous $w>-1$. The results of Table \ref{table1} are then consistent with those obtained by other 
authors.

Although it is usually accepted that the most likely value that makes the existing data 
self-consistent is $w=-1$ (see, for example, Krauss \& Chaboyer \cite{krauss}; Spergel 
et al.\ \cite{spergel}), the fact is that quintessence ($-1/3 > w > -1$) cannot be ruled 
out on observational grounds using currently existing data. Moreover, the assumptions 
of this article do not allow us to rule out phantom energy ($w<-1$). In fact, the 
mean values assumed for $\Omega_{\mathrm{m0}}$, $t_0$, and $t_\mathrm{f}$ give 
$w=-4.95$ for $H_0=72$ km\ s$^{-1}$\ kpc$^{-1}$ and $w=-1.49$ for $H_0$=64 km\ s$^{-1}$\ kpc$^{-1}$. Taking into 
account that the $w$ values inferred are effective values (Section 2), then $w(t)$ must have 
been even smaller, which points strongly to an epoch of the Universe with $w<-1$.

An elementary analysis, varying only one of the parameters  
$\Omega_{\mathrm{m0}}$, $t_{\mathrm{GC}}$, $t_\mathrm{f}$ and $H_0$ at a time, while keeping 
$w=-1$ fixed, gives the following results for the possible options to make the assumed 
\emph{mean values} compatible with a cosmological constant instead of phantom energy: 

\begin{enumerate}
\item Reducing $t_\mathrm{f}$. The age of the Universe derived using (\ref{laecuacion})
and the assumed mean values for $\Omega_{\mathrm{m0}}$ and $H_0$ is 12.7 Gyr, which gives 
no time for globular clusters to form. This possibility alone, then, is not enough.
\item Reducing $t_{\mathrm{GC}}$ to 10.7 Gyr, a possibility that is not excluded by the error 
bars of current age determinations.
\item Reducing $\Omega_{\mathrm{m0}}$ to 0.18, a value barely below the accepted lowest limit of 
0.2 (Reichart et al.\ \cite{reichart}). However del Popolo (\cite{popolo}) suggests that current 
$\Omega_{\mathrm{m0}}$ values obtained from clusters of galaxies are underestimated by 20\%. In 
this case, not only does the possibility of reducing $\Omega_{\mathrm{m0}}$ seem unlikely, but its 
value could even increase, thereby reducing even further the $t_0$ derived from (3) and making 
the need for phantom energy more likely for the current observational data to be consistent.
\item Reducing $H_0$ to 61 km\ s$^{-1}$\ kpc$^{-1}$, a value that seems to be excluded by the HST Key Project
to Measure the Hubble Constant (Freedman et al.\ \cite{freedman}), but that is consistent with 
other determinations using gravitational lenses (Fassnacht et al.\ \cite{fassnacht}), the 
Sunyaev--Zel'dovich effect (Mauskopf et al.\ \cite{mauskopf}) of SNeIa (Parodi et al.\ \cite{parodi}), 
among others.
\item Increasing $\Omega_{\Lambda0}$ to 1.12 to give $\Omega_{\mathrm{total}}=1.45$, incompatible
with current CMBR anisotropy data (e.g., Spergel et al.\ \cite{spergel}; Hu et al.\ \cite{hu}).
\end{enumerate}

Somewhat more rigorously, by parametric derivation of (\ref{laecuacion}) under the integral sign,
a Gaussian error distribution in $w$ centered on $w=-1$, with $w<-0.78$ (95\% c.l. of Spergel 
et al.\ \cite{spergel}), gives $t_0=12.7\pm 0.4$ Gyr at 95\% c.l. for $\Omega_{\mathrm{m0}}=0.33$ 
and $H_0$=72 km\ s$^{-1}$\ kpc$^{-1}$. Taking $t_{\mathrm{GC}}$ at face value leaves at most only 0.2 Gyr for 
globular clusters to form or, equivalently, globular clusters should form around $z=17$ 
or earlier. Considering $\Omega_{\mathrm{m0}}=0.26$ (95\% c.l. of Turner \cite{turner}, and within 
the values obtained by Spergel et al. \cite{spergel}) gives $t_0=13.6\pm0.4$ Gyr at 95\% c.l., 
which leaves 1.1 Gyr, or equivalently $z=5.3$ or earlier for globular clusters to form. 
This possibility could be consistent with the assumption that globular cluster formation can 
happen as early as $z=6$ as claimed by some authors (van den Bergh \cite{bergh}). 

Although obtaining a conclusive result, then, requires more accurate data on 
$\Omega_{\mathrm{m0}}$, $H_0$, $t_{\mathrm{GC}}$ and $t_{\mathrm{f}}$, there 
is prima facie qualitative evidence that the possibilities of reducing the 
current likelihood of phantom energy to favor a cosmological constant as dark energy  
reduces $t_{\mathrm{GC}}$, $H_0$ or both, and possibly advances the epoch of globular cluster 
formation.


\section{The time dependent equation of state}

\subsection{General case}

Except in the case of the cosmological constant, energy densities vary with redshift.
While non-relativistic matter and radiation can be represented by constant $w$, this 
do not generally happen with quintessences and phantom energies, although in these cases
a constant $w$ representing a weighted mean can be taken as an approximation.

As already stated in the introduction, quintessences were invoked to 
avoid the fine-tuning problem posed by the dominance of the cosmological constant 
in recent epochs only. Moreover, quintessences with $w(z)$ might present some additional 
advantages such as string theory compatibility by avoiding the existence of an 
event horizon (Cline \cite{cline}). Many quintessence models 
that differ in the form of the potential have been proposed (see Doran et al.\ 
\cite{doran1} for some potentials used). 

Phantom energies have been represented by Gonz\'alez D\'\i az (\cite{gd}) by means 
of a suitable generalization of the Chaplygin gas model.

In a general case, the differential equation governing the equation of state can be 
derived from the Friedmann equations, 

\begin{equation}
d\ln\rho_{\mathrm{q}}=\frac{3(1+w)}{1+z}dz.
\end{equation}
Then,

\begin{equation}
\rho_{\mathrm{q}}=\rho_{\mathrm{q0}}\exp{3\int^z_0 \frac{1+w}{1+z}dz}\equiv\rho_{\mathrm{q0}}
\exp{Q(z)}
\label{densidad}
\end{equation}
where $Q(z)$ depends on $w(z)$.

With the hypotheses of null curvature and only two equations of state, one of them for 
ordinary matter, (\ref{laecuacion}) can be written as

\begin{equation}
t=\frac{1}{H_0}\int_{z}^{\infty}\frac{\mathrm{d}z}{(1+z)\sqrt{\Omega_{\mathrm{m0}}(1+z)^3+
\Omega_{\mathrm{q0}}\exp{Q}}},
\label{laecuacion2}
\end{equation}
where $\Omega_{\mathrm{q0}}$ is the current adimensional dark energy density. The general
redshift dependence of the adimensional dark energy density is then
$\Omega_{\mathrm{q}}=\Omega_{\mathrm{q0}}\exp{Q}$. From (\ref{laecuacion2}), the age, $t_0$, 
can be derived by setting the lower integral limit to $z=0$.

\subsection{A simple parameterization}

Given the absence of widely accepted and physically indisputable dark energy 
models, a polynomial parameterization has been used by some authors as a simpler 
and pragmatic way to approximate the time dependence of the equation of state. The 
dependence so parameterized could then be fitted to different model predictions. 

For example, Goliath et al. (\cite{goliath}) and Linder (\cite{linder1}) considered 
linear models of the form 

\begin{equation}
w(z)=w_0+w_1z,
\label{lineal1}
\end{equation}
so that

\begin{equation}
\exp{Q}=(1+z)^{3(1+w_0-w_1)}\exp{3w_1z}.
\label{lineal2}
\end{equation}
Linder (\cite{linder2}) also proposed the form 

\begin{equation}
w(z)=w_0+\frac{w_az}{1+z},
\label{inversa1}
\end{equation}
which converges to $w_0+w_a$ at high redshift and gives

\begin{equation}
\exp{Q}=(1+z)^{3(1+w_0+w_a)}\exp{\frac{-3w_az}{1+z}}.
\label{inversa2}
\end{equation}

Both forms were introduced on the basis of reducing the number of parameters to two. 
Assuming that the above expressions can be extrapolated to $z\rightarrow\infty$, an assumption 
otherwise needed to evaluate the age of the universes that they generate, they can be
used to reinforce the need  either to assume phantom energies as dark energies or to revise
the basic cosmological parameters $H_0$, $\Omega_{\mathrm{m0}}$ or $t_{GC}$, $t_f$. 

Although the age of the Universe cannot be used on its own to constrain two parameters, 
additional constraints may be invoked.

A strong constrain is that dark energies cannot dominate at high redshift. Otherwise, the 
expansion would be different and  primordial nucleosynthesis would yield different helium
abundances and the period of structure formation would be affected as well (Doran et al.\
\cite{doran1}). This constraint may be formally represented by 
$\rho_{\mathrm{q}}/\rho_{\mathrm{m}}\rightarrow\ 0$ for $z\rightarrow\infty$, which gives, 
using (\ref{densidad}), (\ref{lineal2}) and (\ref{inversa2}), the conditions $w_1<0$ and
$w_0+w_a<0$ for (\ref{lineal1}) and (\ref{inversa1}), respectively. This strong constraint 
implies that for an equation of state of the form (\ref{lineal1}) phantom energies will 
necessarily be present at high redshift.

A weaker constraint, since it is not based on observational evidence, is that an event 
horizon should not be present since it poses problems for string theories. This is equivalent
to a future non-accelerated universe. Formally, 
$\rho_{\mathrm{q}}/(1+z)^2\rightarrow\mathrm{constant}$ for $z\rightarrow -1$. This condition
gives, using (\ref{densidad}), (\ref{lineal2}) and (\ref{inversa2}), the conditions 
$1+3(w_0-w_1)>0$ and $w_a>0$ for (\ref{lineal1}) and (\ref{inversa1}), respectively.

\begin{figure}
\resizebox{\hsize}{!}{\includegraphics{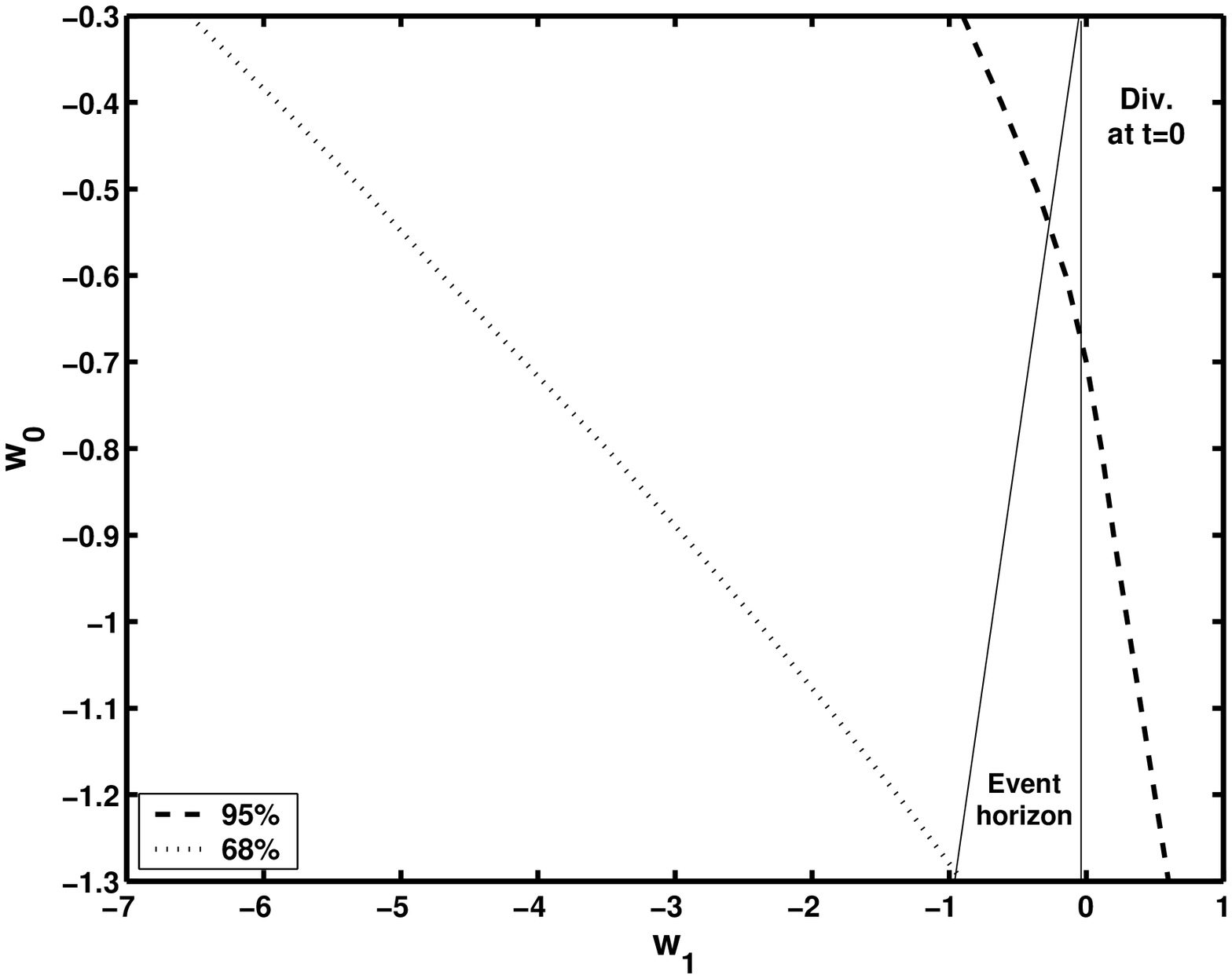}}
\caption{Possible pairs ($w_0$, $w_1$) defining a dark energy equation of
state $w(z)=w_0+w_1z$ that give an age of 12 and 13.4 Gyr  corresponding 
to the minimum age of the Universe at 95\% and 68\% confidence levels, respectively. 
A null-curvature universe with ordinary matter, $\Omega_{\mathrm{m0}}=0.33$ and $t_f=$2 
Gyr has been assumed. The regions $w_1>0$ and $1+3(w_0-w_1)<0$ define respectively the parameter
space were dark energy becomes unacceptably high at high redshifts and an event horizon 
exists.
}
\label{Figure2}
\end{figure}

\begin{figure}
\resizebox{\hsize}{!}{\includegraphics{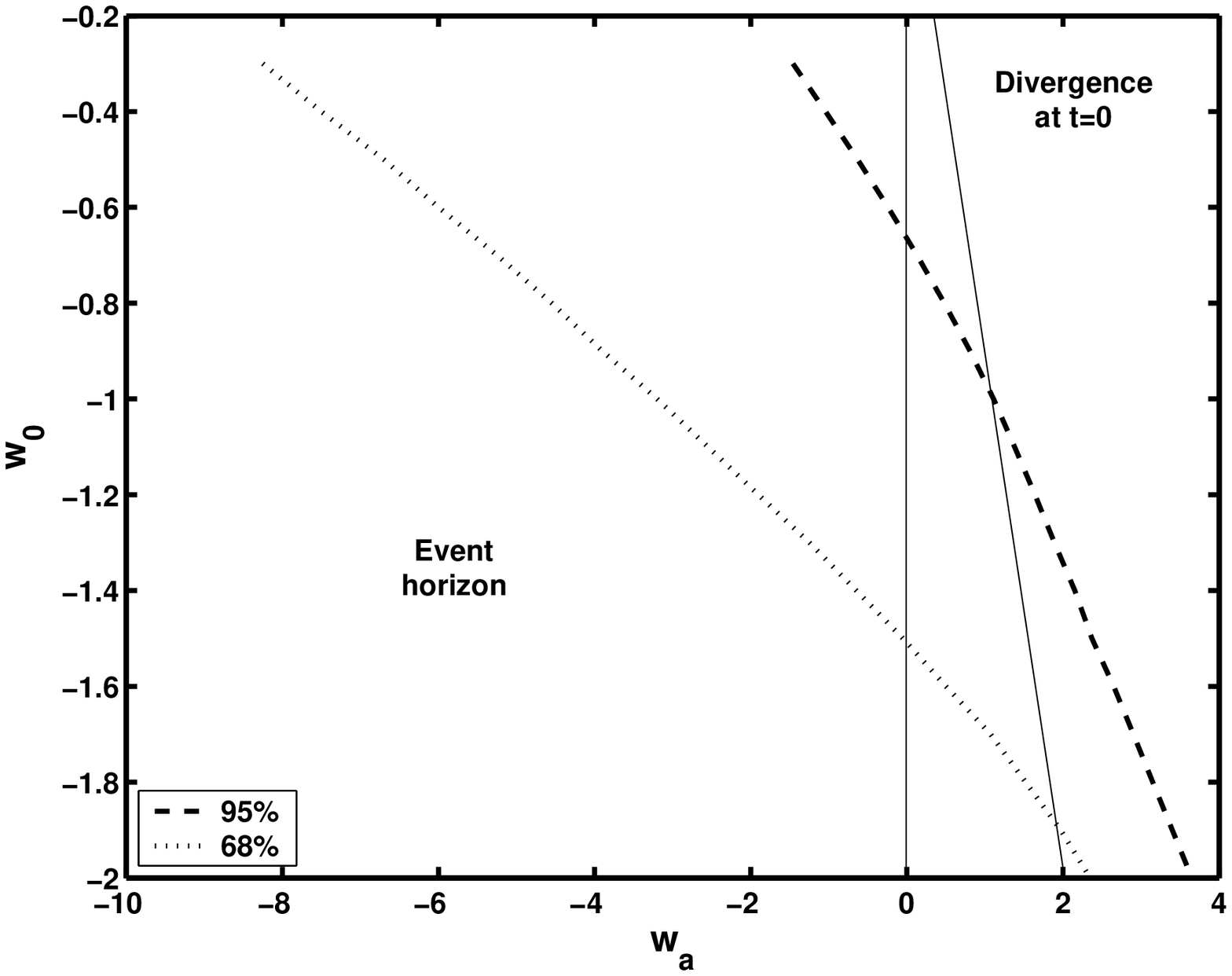}}
\caption{Possible pairs ($w_0$, $w_a$) defining a dark energy equation of
state $w(z)=w_0+w_az/(1+z)$ that give an age of 12 and 13.4 Gyr  corresponding 
to the minimum age of the Universe at 95\% and 68\% confidence levels, respectively. 
A null-curvature universe with ordinary matter, $\Omega_{\mathrm{m0}}=0.33$ and $t_{\rm f}=$2 
Gyr has been assumed. The regions $w_0+w_a>0$ and $w_a<0$ define respectively the parameter
space were dark energy becomes unacceptably high at high redshifts and an event horizon 
exists.
}
\label{Figure3}
\end{figure}

Figures \ref{Figure2} and \ref{Figure3} show the pairs of values for $w_0$ and $w_1$ or
$w_a$, for equations of state of the form (\ref{lineal1}) or (\ref{inversa1}), respectively, 
that give an age of the Universe compatible with the minimum values 
derived from globular clusters at 95\% and 68\% c.l., assuming that globular clusters 
were formed at $z=3$ in a flat universe with $\Omega_{\mathrm{m0}}=0.33$. The regions where
dark energy densities grow unacceptably high at high redshifts are marked. In both cases 
long epochs where $w<-1$ are present, except for some $w_0$, $w_a$ pairs 
corresponding to the lowest 98\% c.l. age values (Figure \ref{Figure3}). Nominal age values 
produce even more negative $w$ values. Hence, the conclusions of the previous section that 
favoring a cosmological constant as dark energy would probably require reducing 
$t_{\mathrm{GC}}$, $H_0$ or both, and possibly advancing the epoch of globular cluster 
formation, are reinforced. 


\section{Age--redshift relations}

The results of Table \ref{table1} show that the upper $w$ limits range between $-0.23$ and $-1.21$. 
The first limit does not accelerate the Universe, since for this effect to happen 
$w<-1/3$ is required. Hence the most extreme upper limit for an accelerated universe may be taken 
as $w=-1/3$. The other extreme may be approximated by $w=-4/3$, since values of $w=n/3$, where 
$n$ is an integer, make obtaining analytical solutions easier. Four 
null-curvature cases will therefore be 
considered: the quintessence limit $w= -1/3$, the quintessence $w= -2/3$, the cosmological 
constant $w= -1$ and the phantom energy $w=-4/3$. These should be considered as weighted
mean approximations to $w$.

\subsection{Quintessence limit ($w = -1/3$)}

As can be deduced from Friedmann equations, this equation of state corresponds to a universe with 
$\ddot a=0$. This is a non-accelerated universe corresponding to the limit between the decelerated 
and accelerated cases. Integrating (\ref{laecuacion}), 

\begin{equation}
t=\frac{1}{H_0}\left[\frac{\sqrt{\Omega_{\mathrm{m0}}(1+z)+\Omega_{\mathrm{w0}}}}
{\Omega_{\mathrm{w0}}(1+z)}+\frac{\Omega_{\mathrm{m0}}}{\Omega_{\mathrm{w0}}^{3/2}}
\ln\left(\sqrt{1+\frac{\Omega_{\mathrm{w0}}}{\Omega_{\mathrm{m0}}(1+z)}}-
\sqrt{\frac{\Omega_{\mathrm{w0}}}{\Omega_{\mathrm{m0}}(1+z)}}\right)\right],
\label{tzq03}
\end{equation}
from which the age of the Universe can be derived:

\begin{equation}
t_0=\frac{1}{H_0}\left[\frac{1}{\Omega_{\mathrm{w0}}}+\frac{\Omega_{\mathrm{m0}}}
{\Omega^{3/2}_{\mathrm{w0}}}\ln\left(\frac{1-\sqrt{\Omega_{\mathrm{w0}}}}
{\sqrt{\Omega_{\mathrm{m0}}}}\right)\right].
\label{t0q03}
\end{equation}

\subsection{Quintessence ($w = -2/3$)}

The integration of (\ref{laecuacion}) give (Gr\"ossner \& Hofreiter \cite{grossner}):

\begin{equation}
t=\frac{1}{H_0}\left\{\frac{2\sqrt{\Omega_{\mathrm{m0}}(1+z)^2+\Omega_{\mathrm{w0}}}}
{\Omega_{\mathrm{w0}}(1+z)+\sqrt{\Omega_{\mathrm{m0}}\Omega_{\mathrm{w0}}(1+z)^3}}
+2\varepsilon\sqrt[4]{\frac{\Omega_{\mathrm{m0}}}{\Omega_{\mathrm{w0}}^3}}
\left[F(\phi,k)/2-E(\phi,k)\right]\right\},
\label{tzq06}
\end{equation}
where $F(\phi,k)$ and $E(\phi,k)$ are elliptic integrals of the first and second 
kinds, respectively, of modulus $k=\sin\pi/4$ and amplitude $\phi$ given by

\begin{equation}
\sin\phi=\frac{2\sqrt[4]{\Omega_{\mathrm{w0}}\Omega_{\mathrm{m0}}}\sqrt{1+z}}
{\sqrt{\Omega_{\mathrm{w0}}}+\sqrt{\Omega_{\mathrm{m0}}}(1+z)}
\end{equation}
for $-\pi/2\leq\phi\leq\pi/2$, with $\varepsilon$ a sign factor

\[\varepsilon=\left\{\begin{array}{ll}
+1 & \mbox{for $0\leq\sqrt{\frac{\Omega_{\mathrm{w0}}}{\Omega_{\mathrm{m0}}}}\leq 1+z$}\\
-1 & \mbox{for $1+z\leq\sqrt{\frac{\Omega_{\mathrm{w0}}}{\Omega_{\mathrm{m0}}}}\leq\infty$}
\end{array}.
\right.\]

The value $1+z=(\Omega_{\mathrm{w0}}/\Omega_{\mathrm{m0}})^{1/2}$ 
corresponds to the coasting point where $\ddot a=0$, which coincides with the
redshift at which the matter and the quintessence energy densities are equal.

Equation (\ref{tzq06}) with $\varepsilon=+1$ is then valid for epochs before the 
coasting point. After the coasting point (as the present epoch would be were 
our universe dominated by this quintessence), it is straightforward to 
demonstrate that to the value of (\ref{tzq06}) for $\varepsilon=-1$ twice 
the value of the second term of (\ref{tzq06}) for $\varepsilon=+1$ has to be 
added. For example, for $z=0$ it is possible to obtain the age of this universe,

\begin{equation}
t_0=\frac{1}{H_0}\left\{\frac{2}{\Omega_{\mathrm{w0}}
+\sqrt{\Omega_{\mathrm{m0}}\Omega_{\mathrm{w0}}}}+2
\sqrt[4]{\frac{\Omega_{\mathrm{m0}}}{\Omega_{\mathrm{w0}}^3}}
\left[K(k)-2E(k)-F(\phi,k)/2+E(\phi,k)\right]
\right\},
\label{t0q06}
\end{equation}
where $K(k)$ and $E(k)$ are the complete elliptic functions of first and
second kinds, respectively, and

\begin{equation}
\sin\phi=\frac{2\sqrt[4]{\Omega_{\mathrm{w0}}\Omega_{\mathrm{m0}}}}
{\sqrt{\Omega_{\mathrm{w0}}}+\sqrt{\Omega_{\mathrm{m0}}}}.
\end{equation}

\subsection{The cosmological constant}

The integration of (\ref{laecuacion}) gives

\begin{equation}
t=\frac{1}{3H_0\sqrt{\Omega_{\Lambda 0}}}\ln\left(
\frac{\sqrt{\Omega_{\mathrm{m0}}(1+z)^3+\Omega_{\Lambda 0}}+\sqrt{\Omega_{\Lambda 0}}}
{\sqrt{\Omega_{\mathrm{m0}}(1+z)^3+\Omega_{\Lambda 0}}-\sqrt{\Omega_{\Lambda 0}}}.
\right)
\label{tz10}
\end{equation}
This provides the known equation for the age of the Universe by taking $z=0$, 

\begin{equation}
t_0=\frac{2}{3H_0\sqrt{\Omega_{\Lambda 0}}}\ln\left(
\frac{1+\sqrt{\Omega_{\Lambda 0}}}
{\sqrt{\Omega_{\mathrm{m0}}}}.
\right)
\label{t010}
\end{equation}
It is interesting to note that, for $\Omega_{\mathrm{m0}}=0.2628$ and 
$\Omega_{\Lambda 0}=0.7372$, 

\begin{equation}
t_0=\frac{1}{H_0}.
\label{cc}
\end{equation}

This generates a curious ``cosmic coincidence'' since these densities are 
within the range currently assumed as the solution for our Universe, in 
particular very close to those assumed by Spergel et al.\ (\cite{spergel}). 
Equation (\ref{cc}) could then be used as a good approximation for the age of the Universe,
were it dominated by a cosmological constant.

Finally, for this equation of state, (\ref{tz10}) can be rewritten to give 
the evolution of the scale factor

\begin{equation}
a=a_0\sqrt[3]{\frac{\Omega_{\mathrm{m0}}}{\Omega_{\Lambda 0}}
\sinh^2\left(\frac{3H_0\sqrt{\Omega_{\Lambda 0}}t}{2}
\right)}.
\label{a}
\end{equation}

\begin{figure}
\resizebox{\hsize}{!}{\includegraphics{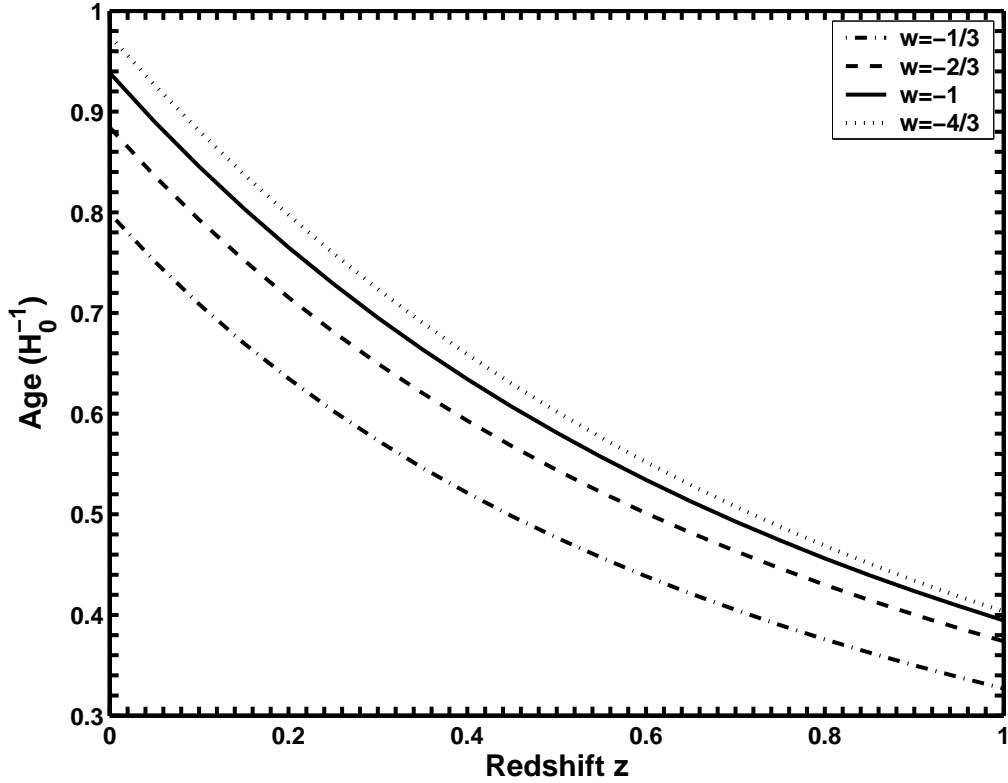}}
\caption{Age--redshift relations in null-curvature universes with ordinary matter
for each of the dark energy equations of state $w$ considered and $\Omega_{\mathrm{m0}}=0.33$.
}
\label{Figure4}
\end{figure}

\begin{figure}
\resizebox{\hsize}{!}{\includegraphics{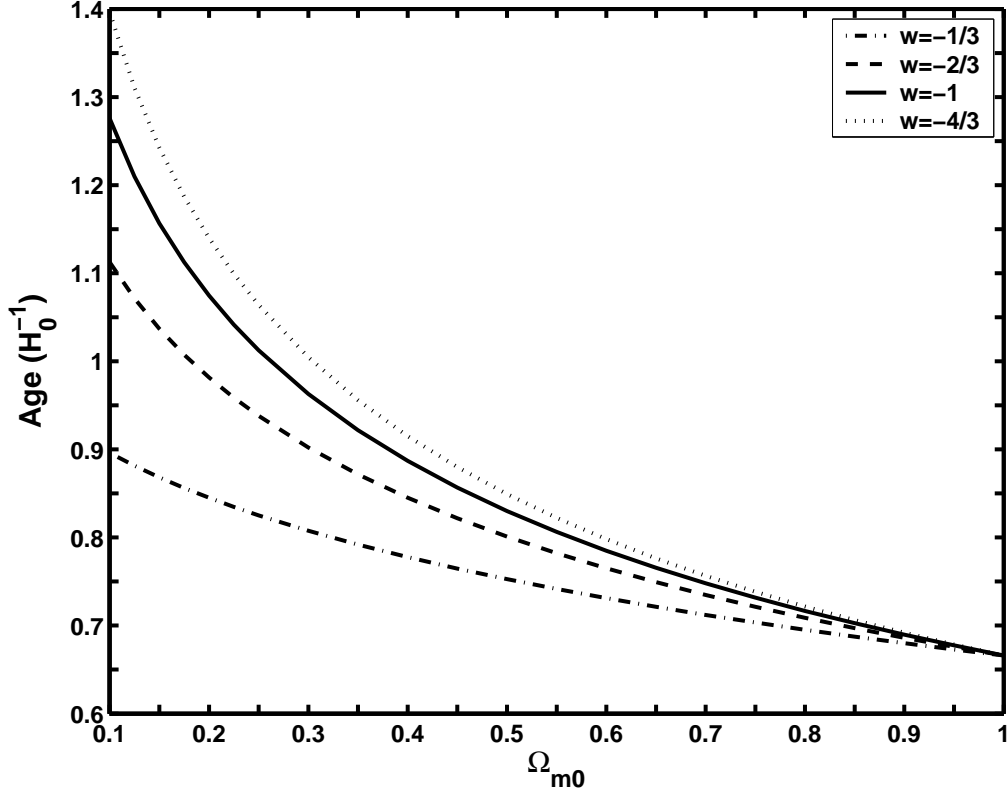}}
\caption{Age of null-curvature universes with ordinary matter versus
$\Omega_{\mathrm{m0}}$ for each of the dark energy equations of state $w$ considered.
}
\label{Figure5}
\end{figure}

\subsection{Phantom energy ($w=-4/3$)}

From the integration of (\ref{laecuacion}) it is possible to obtain (Gr\"ossner 
\& Hofreiter \cite{grossner})

\begin{equation}
t=\frac{\sqrt{2-\sqrt{2}}}{2H_0\sqrt[8]{\Omega_{\mathrm{m0}}\Omega_{\mathrm{w0}}^3}}
\left\{\frac{\sqrt{2}}{2}[F(\phi_1,k)-K(k)]-\varepsilon[F(\phi_2,k)-K(k)]
\right\},
\label{tzp}
\end{equation}
where the modulus $k=\sqrt{2}-1$ and the amplitudes are

\begin{equation}
\sin\phi_1=\frac{\left(\frac{\Omega_{\mathrm{w0}}}{\Omega_{\mathrm{m0}}}\right)
+(1+z)^4-2\sqrt{2}\sqrt[4]{\frac{\Omega_{\mathrm{w0}}}{\Omega_{\mathrm{m0}}}}
(1+z)\left(\sqrt[4]{\frac{\Omega_{\mathrm{w0}}}{\Omega_{\mathrm{m0}}}}+1+z\right)^2}
{\left(\frac{\Omega_{\mathrm{w0}}}{\Omega_{\mathrm{m0}}}\right)
+(1+z)^4+2\sqrt{2}\sqrt[4]{\frac{\Omega_{\mathrm{w0}}}{\Omega_{\mathrm{m0}}}}
(1+z)\left(\sqrt[4]{\frac{\Omega_{\mathrm{w0}}}{\Omega_{\mathrm{m0}}}}+1+z\right)^2}
\label{phi1}
\end{equation}
and
\begin{equation}
\sin\phi_2=\frac{\left(\sqrt[4]{\frac{\Omega_{\mathrm{w0}}}{\Omega_{\mathrm{m0}}}}
-(1+z)\right)^2-\sqrt{2}\sqrt[4]{\frac{\Omega_{\mathrm{w0}}}{\Omega_{\mathrm{m0}}}}(1+z)}
{\left(\sqrt[4]{\frac{\Omega_{\mathrm{w0}}}{\Omega_{\mathrm{m0}}}}
-(1+z)\right)^2+\sqrt{2}\sqrt[4]{\frac{\Omega_{\mathrm{w0}}}{\Omega_{\mathrm{m0}}}}(1+z)},
\label{phi2}
\end{equation}
with $-\pi/2\leq\phi_1,\phi_2\leq\pi/2$ and $\varepsilon$, where

\[\varepsilon=\left\{\begin{array}{ll}
+1 & \mbox{for $0\leq\sqrt[4]{\frac{\Omega_{\mathrm{w0}}}{\Omega_{\mathrm{m0}}}}\leq 1+z$}\\
-1 & \mbox{for $1+z\leq\sqrt[4]{\frac{\Omega_{\mathrm{w0}}}{\Omega_{\mathrm{m0}}}}\leq\infty$}
\end{array}.
\right.
\]
The same comments as for the case $w=-2/3$  apply for the meaning of $\varepsilon$. 
The age of the Universe is then

\begin{equation}
t_0=\frac{\sqrt{2-\sqrt{2}}}{2H_0\sqrt[8]{\Omega_{\mathrm{m0}}\Omega_{\mathrm{w0}}^3}}
\left\{\frac{\sqrt{2}}{2} F(\phi_1,k)+F(\phi_2,k)+K(k)\left(3-\frac{\sqrt{2}}{2}\right)\right\},
\label{t0p}
\end{equation}
with the values of $\phi_1$ and $\phi_2$ given by (\ref{phi1}) and (\ref{phi2}) for $z=0$.

The value $1+z=(\Omega_{\mathrm{w0}}/\Omega_{\mathrm{m0}})^{1/4}$ 
corresponds to the redshift at which the energy densities of dark and conventional matter are equal.
The coasting point where $\ddot a=0$ is given by 
$1+z=(3\Omega_{\mathrm{w0}}/\Omega_{\mathrm{m0}})^{1/4}$.

Again, (\ref{cc}) provides a good estimate for the age of the Universe when $w=-4/3$ for 
$\Omega_{\mathrm{m0}}=0.3054$ and $\Omega_{\mathrm{w0}}=0.6946$.

Figure \ref{Figure4} gives the age--redshift relation for each equation of state for $\Omega_{\mathrm{m0}}=0.33$ 
and $\Omega_{\mathrm{w0}}=0.67$. Figure \ref{Figure5} gives the age of the Universe for each equation of state 
versus $\Omega_{\mathrm{m0}}$. Table \ref{table2} gives the age of the Universe derived for each 
equation of state considered. Abramowitz \& Stegun (\cite{abramowitz}) has been used 
for the evaluation of elliptical integrals.

\begin{table}
\begin{center}
\caption[]{Values of $H_0t_0$ for the different equations of state of dark energy and different 
$\Omega_{\mathrm{m0}}$ for $\epsilon=0$}
\label{table2}
\begin{tabular}{ccccc}
\hline
\noalign{\smallskip}
$w$ &  $\Omega_{\mathrm{m0}}=0.24$ & $\Omega_{\mathrm{m0}}=0.27$ & 
$\Omega_{\mathrm{m0}}=0.30$ & $\Omega_{\mathrm{m0}}=0.33$\\
\noalign{\smallskip}
\hline
\noalign{\smallskip}
$-$1/3 & 0.8302 & 0.8191 & 0.8088 & 0.7991 \\
$-$2/3 & 0.9473 & 0.9241 & 0.9033 & 0.8844 \\
$-$1   & 1.0251 & 0.9927 & 0.9641 & 0.9386 \\
$-$4/3 & 1.0790 & 1.0399 & 1.0057 & 0.9754 \\
\noalign{\smallskip}
\hline
\end{tabular}
\end{center}
\end{table}


\section{Summary}

In the most accepted cosmological model the Universe is currently accelerated 
owing to a for{\bf m} of dark energy of unknown origin, as found via distance--redshift fits 
to high redshift SNeIa. This energy complements the fraction of ordinary matter so that the 
curvature is null, according to the observed CMBR anisotropy and the inflation paradigm. 
Also, the dark energy increases the age of the Universe for the same values of 
$\Omega_{\mathrm{m0}}$ and $H_0$ with respect to a universe composed of non-relativistic 
matter only, thus alleviating possible age conflicts.

In this article a recent determination of $\Omega_{\mathrm{m0}}$, $t_0$ and $H_0$ has been 
selected from the literature with the criterion of providing a set of cosmological parameters 
as model-independent as possible. A simple analysis shows that the weighted mean equation 
of state of dark energy has upper values in the range $-0.23 > w > -1.21$. Hence, not 
only a cosmological constant but also quintessence or phantom energy are also viable dark 
energy candidates from an observational point of view. In fact, the mean values of 
$\Omega_{\mathrm{m0}}$, $t_0$ and $H_0$ favor phantom energy over other alternatives. 
Moreover, a time-dependent $w$ favors even more the existence of epochs 
with $w<-1$. It is qualitatively shown that to reduce the likelihood of phantom energy 
to favor a cosmological constant as dark energy, smaller globular cluster ages and/or 
smaller $H_0$ values with respect to those currently determined  are required. 
Pushing back globular cluster formation to earlier epochs cannot be ruled out, but only together 
with lower values of $t_{\mathrm{GC}}$ and/or $H_0$. Otherwise, an age conflict might still be 
present.

The range of upper $w$ limits inferred is approximately covered by taking $w=-n/3$ for 
$n=1,2,3,4$. For each of these constant values of $w$, which include quintessence, the 
cosmological constant and phantom energy, analytical age--redshift solutions for Euclidean 
universes have been deduced and analytical age expressions derived. Some of these analytical 
solutions have not been derived before. Furthermore, no analytical expressions seem to 
be available in the literature for the range of the constant equations of state 
considered. Analytical equations are more precise, faster and readier to use than numerical 
calculations when evaluating the age of the Universe. Moreover, the explicit dependence on 
$\Omega_{\mathrm{m0}}$ and $\Omega_{\mathrm{w0}}$ renders their study easier. Finally, 
curious ``cosmic coincidences'' make 1/$H_0$ a good approximation for the age of the 
Universe for the currently accepted $\Omega_{\mathrm{m0}}$ values assuming null curvature 
and a cosmological constant or a phantom energy as dark energies.

\begin{acknowledgements}
This work has been supported by the Spanish \emph{Plan Nacional de Astronom\'\i a y 
Astrof\'\i sica\/} under grant AYA2002-01379. This research has made use of NASA's Astrophysics 
Data System.
\end{acknowledgements}

\end{document}